# High Density Fabrication Process for Single Flux Quantum Circuits

D. Yohannes, M. Renzullo, J. Vivalda, A. C. Jacobs, M. Yu, J. Walter, A. F. Kirichenko,
I. V. Vernik and O. A. Mukhanov*

SEEQC, Inc., 150 Clearbrook Road, Elmsford, New York, 10523, USA

*Corresponding author: omukhanov@seeqc.com

**Abstract:** We implemented, optimized and fully tested over multiple runs a superconducting Josephson junction fabrication process tailored for the integrated digital circuits that are used for control and readout of superconducting qubits operating at millikelvin temperatures. This process was optimized for highly energy efficient single flux quantum (ERSFQ) circuits with the critical currents reduced by factor of ~10 as compared to those operated at 4.2 K. Specifically, it implemented Josephson junctions with 10 µA unit critical current fabricated with a 10 µA/µm$^2$ critical current density. In order to circumvent the substantial size increase of the SFQ circuit inductors, we employed a NbN high kinetic inductance layer (HKIL) with a 8.5 pH/sq sheet inductance. Similarly, to maintain the small size of junction resistive shunts, we used a non-superconducting PdAu alloy with a 4.0 ohm/sq sheet resistance. For integration with quantum circuits in a multi-chip module, 5 and 10 µm height bump processes were also optimized. To keep the fabrication process in check, we developed and thoroughly tested a comprehensive Process Control Monitor chip set.

The low integration scale of superconducting Josephson junction circuitry has always been a concern, especially for Single Flux Quantum circuits, where the product of the junction critical current $I_c$ and circuit inductance $L$ should be in the range from 0.1 to 1.0 $\Phi_0$, where $\Phi_0$ is flux quantum. This becomes an even bigger problem when SFQ circuits are scaled to work in the millikelvin environment necessary for various applications where they are integrated with other technologies operating at this temperature, including quantum circuits (qubit arrays) [1,2], sensor arrays [3], etc. The delicate nature of qubits and the limited cooling capacity of the cryogenic platforms call for the minimization of the power dissipation, while still retaining the speed advantage of SFQ technology. The straightforward way to achieve this is to scale down the Josephson junction critical currents $I_c$ as the energy dissipated during SFQ switching at JJ is $E_{SFQ} \sim I_c \Phi_0$. The reduction of critical currents is possible due to the lower thermal noise at mK. However, simply scaling down the critical current of a Josephson junction causes a proportional increase in circuit inductances and Josephson junction shunt resistances necessary to retain circuit functionality. In addition, the bias inductors of ERSFQ circuits [4] grow so large that inductive coils become necessary. As a result, the ERSFQ circuit area substantially increases, hindering the prospects of scaling to greater circuit complexities. To tackle these issues, new materials should be employed. A superconducting high kinetic inductance layer (HKIL) can address the inductors' physical size issue. In this superconducting film, the London penetration depth is larger than the film thickness, leading to substantially higher kinetic inductance. The first successful application of the HKIL was done for the implementation of large bias inductors in the ERSFQ 8 bit arithmetic logic unit (ALU) [5] using MIT-LL process with a $Mo_2N$ HKIL [21], with the circuit cell inductors still implemented using conventional Nb layers. Very recently, MIT-LL proposed using bilayers of $Mo_2N$/Nb and NbN/Nb with higher sheet inductances of 8 pH/sq and 3 pH/sq, respectively [6], which can be used for both large and small circuit inductances. In this paper, we report the development, characterization and successful use of the area-efficient SFQuClass fabrication process for ERSFQ circuits, optimized for integration with qubit chips operating in a 20 mK cryogenic environment. This process enables higher density, more energy efficient ERSFQ circuits by employing the HKIL for both bias and circuit inductors, as well as using high resistivity junction shunts and scaling the minimal critical current down to 10 μA.

The new fabrication process for SFQ integrated circuits capable of operating at millikelvin temperatures and integrated with qubit chips (SFQuClass circuits [2]) was derived from the standard HYPRES process with all Nb layers [7]. The new process, SFQ-C5SL, has five superconducting layers – four Nb layers and an additional NbN layer. The superconducting layers are separated by a bi-layer interlayer dielectric of silicon nitride and silicon dioxide to take advantage of their chemical and material properties. The cross section of the process is shown in Fig. 1; the target parameters of the layers as they appear in the processing flow are given in Table 1.

We divide the integrated circuit fabrication into a sequential execution of unit processes. Each unit fabricates a layer of either a superconducting metal or an interlayer dielectric. A typical unit process starts with the deposition of the respective material layer with the required thickness onto a wafer, followed by photolithography, and then patterning of the layer and removal of the residual photoresist. As described below, we use chemical-mechanical polishing (CMP) and anodization steps for selected layers. In process characterization, we performed measurements of thickness, sheet resistance, and dielectric integrity using a profilometer, ellipsometer, four-point prober, wafer prober, scanning electron microscope (SEM) and atomic force microscope (AFM). All superconductor materials are deposited by DC magnetron sputtering in high vacuum chambers. The bilayer $SiN/SiO_2$ is deposited in two separate systems using plasma enhanced chemical vapor deposition (PECVD). Most of the photolithography is done on a 5x reduction deep-UV stepper and some on an i-line laser writer with 0.6 um resolution. The patterned metal or dielectric surfaces are dry-etched in either reactive ion etcher (RIE) or inductively coupled plasma (ICP) etcher. All processes are performed at less than 180 °C process temperature so the residual photoresist can be easily removed with standard solvents.

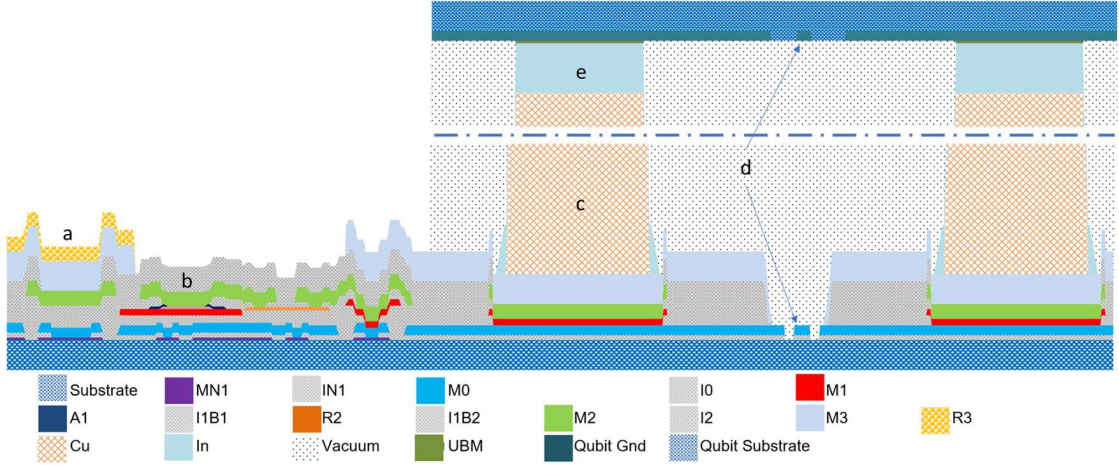

*Figure 1. A cross-section of the multi-chip module consisting of classical SFQ chip (bottom) and a quantum chip (top): a – contact pad, b - shunted JJ, c – Al or Cu bump core, d – qubit-SFQ coupling, e- indium bump cap. Dashed line indicates a break line in the bump hight.*

| Layer | Thickness [nm] | Material, Description |
| --- | --- | --- |
| Substrate | 650000 | High resistivity silicon |
| MN1 | 40 | NbN, High kinetic inductance layer 8.5 pH/sq |
| IN1 | 25/75 | $SiN_x/SiO_2$ |
| M0 | 200 | Nb |
| I0 | 150 | $SiO_2$ |
| J1 | 50 | Nb, JJ-top electrode |
| A1 | 40 | AlOx/NbOx, Anodized layer |
| M1 | 135 | Nb, JJ-bottom electrode |
| R2 | 40 | PdAu |
| I1 | 100/100 | $SiN/SiO_2$ |
| M2 | 300 | Nb |
| I2 | 250/250 | $SiN/SiO_2$ |
| M3 | 600 | Nb |
| I3 | -1100 | $SiN/SiO_2$ etch down to Si |
| R3 | 40/100/200 | Mo/Pd/Au |
| Bump_1 | 40/50/10000/50 | Mo/Ti/Cu/Ti, Hard stop bump material (Cu or Al) |
| Bump_2 | 200/50/1000 | NbN/Ti/In, Indium coat |

*Table 1. SFQ-C5SL process layers.*

*High kinetic inductance layer (MN1, IN1)*: This is the first superconducting layer for SFQ-C5SL process. It is a 40 nm thick NbN film deposited directly on a high resistivity 150 mm silicon wafer. The target sheet inductance for this layer is 8.5 pH/sq. This is significantly higher than the 0.4 pH/sq sheet inductance of the most frequently used

Nb layer placed between two ground planes. This allows a ~20x reduction of the length of the inductor, which is critical for the circuit area. The patterning, etching and passivation of this HKIL requires a special development that results in a more controllable process. The particular problem is in etching: most chemistries for etching NbN etch Si even faster, risking a deep overetch into the silicon substate and resulting in edge coverage issues and potential shorts.

*Ground plane (M0, I0):* *M0* is the second superconducting layer of the process and is deposited after the patterning and etching of the *IN1* vias to the *MN1* layer. The interlayer dielectric between *M0* and *MN1* is a 100 nm thick SiN/SiO$_2$ bilayer. *M0* is a 200 nm thick layer to which all ERSFQ circuits are grounded. It also has intricate patterns, such as moats, to mitigate flux trapping [8-10]. The coplanar waveguides (CPW) and resonators are often made in this layer. The CMP is performed on the dielectric layer after patterning of MN1, IN1 and M0 layers. After the CMP, the I0 vias are patterned and etched. This enables a planarized flat surface before the deposition of the trilayer. The CMP process is carefully monitored and kept at a 20 nm standard deviation in thickness. This CMP step is necessary for preventing the topography from translating to the upper layers and potentially inducing shorts between layers due to edge coverage issues.

*Josephson Junction (JJ, A1, M1):* The next superconducting layer to be deposited is the base electrode of the Josephson Junctions (*M1*), a 150 nm Nb layer with a 10 nm Al layer underneath. The thin Al layer under the Nb layer serves as a hydrogen diffusion barrier [11]. This is followed by *in situ* deposition of another 10 nm Al layer. After that, the wafer is transferred to an oxidation chamber to grow the tunnel barrier in a controlled manner by oxidizing the aluminum at specific pressure and time to achieve critical current density of 10 $\mu A/\mu m^2$. Then, the wafer is placed back into the sputter chamber without breaking the vacuum to deposit 50 nm of Nb (the counter electrode of the junction). After that, the wafer is passivated by a 25 nm low-stress PECVD SiN layer deposited at 150 °C before patterning the *J1* layer. After patterning *J1*, the whole wafer is anodized to grow oxide around the JJs, providing protection from accidental shorts during the *IN1* via etching. Such a protective layer is only necessary around a JJ and is removed everywhere else. This is implemented by the *A1* mask patterning and dry etching everywhere except the area around a JJ. Finally, the *M1* (a third superconducting layer) is patterned.

*Shunt Resistors (I1-1/R2/I1-2):* In ERSFQ, resistors are needed for shunting Josephson junctions, impedance matching, etc. Our current material of choice is a Pd/Au mixture, since the resistors have to stay non-superconducting at millikelvin temperatures. We start with depositing a 100 nm interlayer dielectric (*I1-1*) of PECVD-SiN. Then we deposit and pattern a photoresist with the R2 mask, followed by the resistor film evaporation deposition, with the target sheet resistance of 4.0 ohm/sq at 4.2 K. Afterwards, the resistors are patterned using a lift-off process. The next step is a deposition of the interlayer dielectric (*I1-2*) of PECVD-SiO$_2$. The *I1* patterning by high-selective etching is used to open vias to the junctions' counter electrodes, the base electrodes, and to the resistors.

*Superconducting wiring layer (M2/I2):* The fourth superconducting layer is the M2 layer with a 300 nm thickness. This layer is used for wiring and for connecting resistors and junctions. Some circuit inductors are also defined in this layer.

*Sky plane (M3/I3):* The fifth superconducting layer is *M3* – the top ground layer that is also referred to as the skyplane. It shields most of the area of the carrier facing the qubit chip. The areas that interact with qubits, like the feedthrough and control tips, are etched with *I3* pattern to remove all exposed dielectric. This way we can decrease the participation ratio of the dielectric and preserve high qubit performance [12] in the MCM configuration.

*Contact pad (R3):* The contact pads need to be coated with non-oxidizing metal to ensure good contact during the testing. This is done with a lift-off process. A 70 nm molybdenum film is deposited by sputtering, followed by the evaporation of a 100/200 nm Pd/Au bilayer. The molybdenum layer acts as a hydrogen diffusion barrier [11].

*Bumps (Al/In):* The final step of the process is the deposition of the Bump layer. The bumps are fabricated in two steps. First, depending on a qubit design, 5 or 10 μm high aluminum or copper bump cores are deposited by evaporation. Then after the first lift-off, the second is patterned, and a 1.0 μm thick indium layer is evaporated, preceded by a coating with a 200 nm NbN film. The NbN film acts as a diffusion barrier between the Al (or Cu) core and In layer, preventing a potential formation of brittle intermetallic compounds. The bilayer Al-In design of bumps ensures better control of the interchip spacing and the planarity. The core provides the necessary strength to hold the spacing, and the thin In provides the necessary adhesion while maintaining the planarity to < 0.5 um bump-to-bump uniformity across the chip.

In order to provide a comprehensive process control, we designed a reticle comprising four 5 mm by 5 mm process diagnostic chips: three with conventional process control monitors (PCMs) and one with representative ERSFQ circuits and cells. This *digital* PCM chip with several ERSFQ circuits of various complexity provides additional information highly relevant to our SFQ circuits complementing the data obtained by measuring typical *analog* PCMs [6, 7] (such as Josephson junctions and arrays as well as structures to test sheet resistances, sheet inductances, vias, in-layer gaps shorts, interlayer shorts, etc.). The integrated diagnostic reticle with analog and digital PCMs is placed in the center and four corners of all project wafers to characterize and monitor the fabrication process quality, uniformity, and reproducibility. The analog PCM structures include:

- Several arrays of nominally shunted JJs of varied sizes for monitoring the critical current density and the process and lithography bias (also called "missing radius" (*dr)*).
- A single unshunted JJ for monitoring trilayer quality parameters, like Nb gap voltage ($V_g$), sub-gap resistance parameter ($V_m$), etc.
- A test structure for measuring *R2* sheet resistance comprising a chain of resistors of varying widths and lengths.
- SQUID-based inductance test structures [13] to extract sheet inductance of the NbN and Nb films along with the overetch and the fringing factor. This allows characterization of uniformity of interlayer dielectric (ILD) thickness over the wafer.

- A mesh in *MN1* covered by the identical, 90° rotated mesh in *M0* layer, covered by 90° rotated mesh in *M1*, covered by 90° rotated mesh in *M2*, and finally covered by 90° rotated mesh in *M3* to assess quality of planarized and non-planarized insulation between these layers by testing for interlayer shorts.
- Arrays of vias between *MN1-M0-M1* and all neighboring metal layers (*MN1-M1, M0-M1*, etc.) to measure their critical currents.
- Long gap test structures in all metal layers with gap sizes ranging from 0.5 μm to 1.6 μm by testing for shorts.

The analog PCMs are routinely tested at room temperature and 4.2 K, with some structures tested at both temperatures and others only at 4.2 K. We test unshunted JJs to monitor the Nb gap voltage ($V_g = 2.65 \pm 0.05$ mV) and sub-gap resistance parameter ($V_m$ in excess of 30 mV). The critical current of shunted JJs can be well fitted with the following equation $I_c = \pi \cdot j_c \cdot (r-dr)^2$, where $r$ is the designed junction radius and $dr$ is the proc.ess/lithography bias, $j_c$ is the critical current density. Here we use shunted JJs because they are less susceptible to noise. Based on the results of these measurements, oxidation parameters for the trilayer fabrication were calibrated and on-mask compensation for the overetch was accounted for. A variation in $j_c$ of 1σ less than 3% over a 150 mm wafer was measured.

Since HKIL MN1 layer is used not only in the biasing but also within the ERSFQ circuits themselves, the tight control of the run-to-run variations and wafer spread of the *MN1* sheet inductance becomes imperative. We calculated *MN1* sheet inductance $L_{sq}$ and overetch $dw$ by computing the linear regression $L_s = L_{sq}*(w+dw)$ for the inductances of six different width inductors ranging from 1.0 μm to 2.2 μm. From wafer to wafer, we regularly hit sheet inductance for *MN1* layer within ±10% of target value of 8.5 pH/sq with a 1σ spread over a 150 mm wafer of 8% or less. Figure 2 shows relevant *MN1* inductance data collected over 34 wafers fabricated over two years. Results outside of ±10% boundaries, or somewhat higher 1σ spread, are obtained when the regression was done with one or more absent data points for some inductances. This was traced back to a test malfunction due to a contact issue. While the most reliable *MN1* sheet inductance is measured at 4.2 K, we are developing a method to characterize it at room temperature as well. The inductance is not expected to change significantly at the intended temperature of operation according to the theoretical considerations [14] and measurements [15] showing nearly constant kinetic inductance of NbN films at $T \ll T_c$. The measured dependence in Figure 3 suggests that in the fabrication process described here, we should target room temperature sheet resistance in the range of 63 Ω/sq to 70 Ω/sq.

The digital PCM chip comprises seven ERSFQ circuits. They are, in order of increasing complexity (in terms of JJ count and function), as follows: two chains of DC/SFQ-JTL-SFQ/DC, where JTL is a Josephson transmission line, (one realized with HKIL inductances and the other without them), a D flip-flop (DFF) [16,17], a T flip-flop (TFF) [16,17], a frequency divider by 4 comprising two serially connected TFFs, a D2 flip flop [18], and an NDRO switch comprising an NDRO cell [16,17] with the synchronized set/reset inputs.

To show the effect of the HKIL-implemented ERSFQ circuit inductors, Figure 4 compares two very similar DC/SFQ-JTL-SFQ/DC circuits. The circuit in Figure 4b is realized employing cell inductors in HKIL *MN1* NbN layer, while the circuit in in Figure 4a is designed with conventional SEEQC fabrication process [7] with all inductors in *M1* and *M2* Nb layers. The most obvious difference is that the circuit in Figure 4b is more compact and occupies ~2.5 times smaller area. The bounding box in Figure 4a and strip in Figure 4b (highlighted by an arrow) show identical value inductances realized in M2 Nb and MN1 NbN layer, respectively. The *MN1* circuit inductors in Figure 4b are not visible because of *IN1* layer planarization. Layout design using *MN1* circuit inductors is more straightforward by using simple strip inductors, avoiding most parasitic inductances that are in the low-inductive *M1* and *M2* layers. Additionally, for circuits with low critical currents, the effect of such parasitic inductances is significantly diminished. Similarly, the parasitic mutual inductances become negligible since kinetic inductances do not generate magnetic coupling while geometric inductances are very small.

All ERSFQ circuits from the digital PCM chip were designed and optimized for at least $\pm25\%$ bias margins. Note that all PCM circuits have one shared bias line for the cell and the input/output monitors. As with the analog PCM, the digital PCM chips from the center and four corners of all fabricated wafers are routinely tested. Figure 5 shows a typical error rate dependence for the NDRO switch measured at 4.2 K. The NDRO switch is the most complex circuit from the digital PCM chip, both in logic operation and JJ count. Together with all input/outputs, it consists of 119 JJs. The error rate in Figure 5 is measured by testing the switch 100 times at each bias current and is used as a metric of quality and reproducibility of the fabrication process. The NDRO switch in Figure 5 shows $\pm12\%$ bias margins with ~0% error rate in its operational regime. Note that the circuit is designed to operate at mK temperatures with a smallest JJ critical current of 10 µA. Its error rate at 4.2 K is strongly affected by the thermally activated switching of JJs.

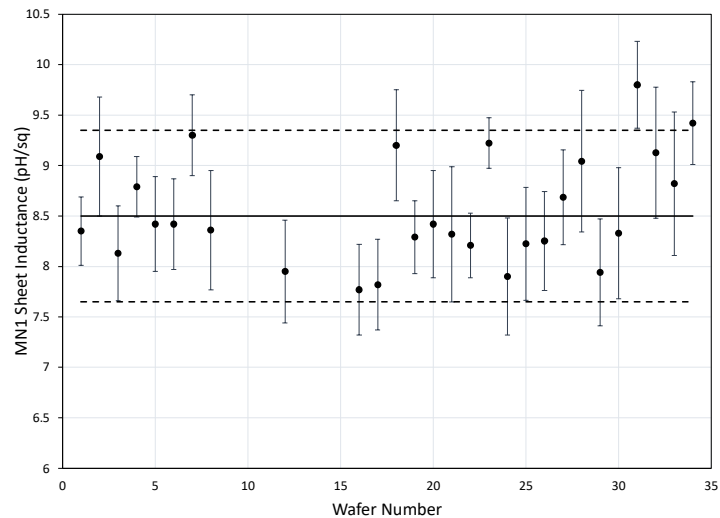

*Figure 2. MN1 HKIL sheet inductance measured at 4.2 K for the wafers fabricated over a 2-year period. The horizontal lines denote 8.5 pH/sq target value (solid line) and $\pm10\%$ boundaries (dash lines). Error bars show $1\sigma$ spread over a 150 mm wafer.*

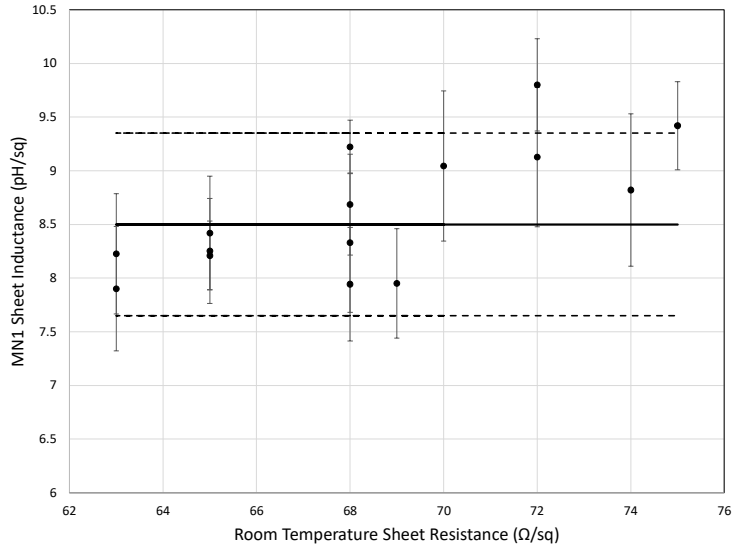

*Figure 3. MN1 HKIL sheet inductance at 4.2 K vs room temperature sheet resistance. The horizontal lines denote 8.5 pH/sq target value (solid line) and ±10% boundaries (dash lines). Error bars show 1σ spread over a 150 mm wafer.*

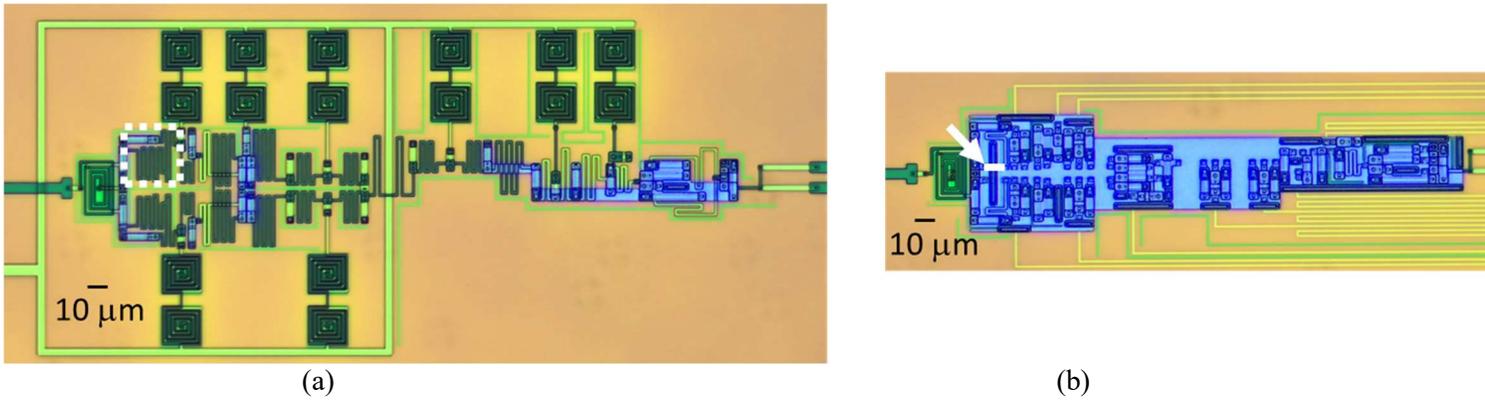

(a)　　　　　　　　　　　　　　　　　　　　　　　　　　(b)

*Figure 4. DC/SFQ-JTL-SFQ/DC circuits realized without (a) and with (b) the use of MN1 HKIL layer. Bounding box in (a) and strip indicated by arrow in (b) are inductances equal in value.*

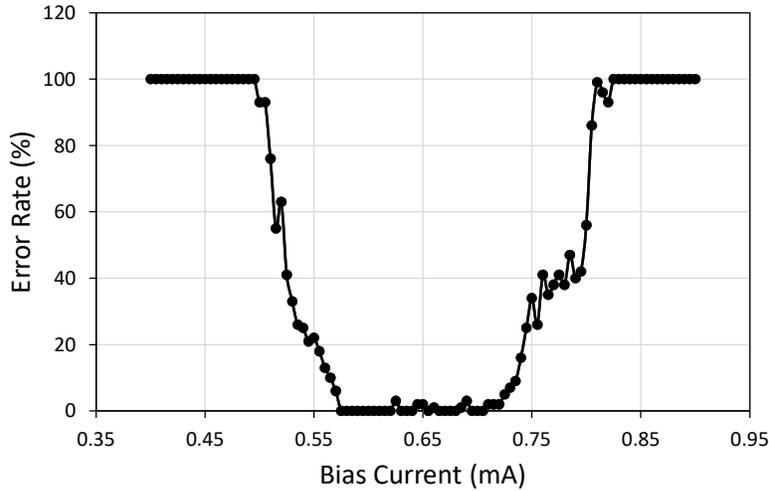

*Figure 5. Measured error rate dependence vs. bias current for ERSFQ NDRO based switch measured at 4.2 K tested 100 times at each bias current.*

We have developed, verified and optimized over many runs a fabrication process SFQ-C5SL targeted for the implementation of high-density energy-efficient SFQ circuits operating in a millikelvin environment in close proximity to quantum circuits. The key advantage of this process is the use of a high kinetic inductance layer for the compact implementation of the bias and circuit inductors, while also diminishing the influence of parasitic inductances. To maintain the small size of junction resistive shunts, we introduced a PdAu mixture with higher sheet resistance. Ultimately the junction shunt resistors can be eliminated, e.g., by using different barrier material [19]. The challenge of achieving high reproducibility, low spread across the wafer and from run-to-run has been addressed by optimizing the fabrication recipes, design rules, and layout verification procedures. In order to keep the fabrication process within specifications, we introduced a comprehensive set of diagnostic chips, including the digital ERSFQ circuit set, capable of providing the most relevant feedback to integrated circuit processing and design.

Since kinetic inductances are not suitable for the implementation of magnetic transformers needed for ac-biased superconducting circuits, they can instead be fabricated using the conventional Nb layers *M1/M2*, although with the inevitable increase in circuit area. Potentially, one could implement very compact transformers using layered magnetic structures similar to those described in [20].

**References:**

[1] R. McDermott, M. G. Vavilov, B. L. T. Plourde *et al*., "Quantum-Classical Interface Based on Single Flux Quantum Digital Logic," *Quantum Sci. Technol.* **3**, 024004 (2018).

[2] O. Mukhanov, A. Kirichenko, C. Howington *et al.,* "Scalable Quantum Computing Infrastructure Based on Superconducting Electronics," *2019 IEEE International Electron Devices Meeting (IEDM)*, San Francisco, CA, USA, 2019, doi: 10.1109/IEDM19573.2019.8993634.

[3] A. Suzuki, N. Cothard, A. T. Lee *et al.,* "Commercially Fabricated Antenna-Coupled Transition Edge Sensor Bolometer Detectors for Next-Generation Cosmic Microwave Background Polarimetry Experiment," *J. Low Temp Phys*. **199**, 1158–1166 (2020).

[4] D. E. Kirichenko, S. Sarwana and A. F. Kirichenko, "Zero Static Power Dissipation Biasing of RSFQ Circuits," *IEEE Trans. Appl. Supercond.* **21**, 776-779 (June 2011).

[5] A. F. Kirichenko, I. V. Vernik, M. Y. Kamkar *et al*., "ERSFQ 8-bit Parallel Arithmetic Logic Unit," *IEEE Trans. Appl. Supercond.* **29**, 1302407 (2019). .

[6] S. K. Tolpygo, J.L. Mallek, V. Bolkhovsky *et al*., "Progress Toward Superconductor Electronics Fabrication Process with Planarized NbN and NbN/Nb Layers," *IEEE Trans. Appl. Supercond*, doi: 10.1109/TASC.2023.3246430.

[7] D. T. Yohannes, R. T. Hunt, J. A. Vivalda *et al*., "Planarized, Extendible, Multilayer Fabrication Process for Superconducting Electronics," *IEEE Trans. Appl. Supercond*. **25**, 1100405 (2014).

[8] R. P. Robertazzi, I. Siddiqi, and O. A. Mukhanov, "Flux Trapping Experiments in Single Flux Quantum Shift Registers," *IEEE Trans. Appl. Supercond.* **7**, 3164-3167 (1997).

[9] S. Narayana, Y. A. Polyakov, and V. K. Semenov, "Evaluation of flux trapping in superconducting circuits," *IEEE Trans. Appl. Supercond*. **19**, 640–643 (2009).

[10] C. J. Fourie, K. Jackman, "Experimental Verification of Moat Design and Flux Trapping Analysis," *IEEE Trans. Appl. Supercond.* **31**, 1300507 (2021).

[11] S. K. Tolpygo, D. Amparo, R. T. Hunt, J. A. Vivalda and D. T. Yohannes, "Diffusion Stop-Layers for Superconducting Integrated Circuits and Qubits with Nb-Based Josephson Junctions," *IEEE Trans. Appl. Supercond.* **21**, 119-125 (2011).

[12] C. Wang, C. Axline, Y. Y. Gao, *et al*., "Surface participation and dielectric loss in superconducting qubits," *Appl. Phys. Lett.* **107**, 162601 (2015).

[13] D. Yohannes, S. Sarwana, S Tolpygo *et al*., "Characterization of HYPRES' 4.5 kA/cm$^2$ & 8 kA/cm$^2$ Nb/AlO$_x$/Nb Fabrication Processes," *IEEE Trans. Appl. Supercond*. **15**, 90-93 (2005).

[14] S. K. Tolpygo, E. B. Golden, T. J. Weir *et al*., "Self-and mutual inductance of NbN and bilayer NbN/Nb inductors in a planarized fabrication process with Nb ground planes," *IEEE Trans. Appl. Supercond*. doi: 10.1109/TASC.2023.3244772.

[15] A. J. Annunziata, D. F. Santavicca, L. Frunzio, G. Catelani, M. J. Rooks, A. Frydamn, D. E. Prober, "Tunable superconducting nanoinductors," *Nanotechnology* **21**, 445202 (2010).

[16] K. K. Likharev and V. K. Semenov, "RSFQ Logic/Memory Family: A New Josephson-junction Technology for Sub-Terahertz-Clock-Frequency Digital Systems," *IEEE Trans. Appl. Supercond*. **1**, 3–28 (1991).

[17] S. V. Polonsky, V. K. Semenov, P. Bunyk *et al*., "New RSFQ Circuits," *IEEE Trans. Appl. Supercond.* **3**, 2566-2577 (1993).

[18] S. V. Polonsky, V. K. Semenov, A. F. Kirichenko, "Single Flux Quantum B Flip-Flop and Its Possible Applications," *IEEE Trans. Appl. Supercond*. **4**, 9-16 (1994).

[19] D. Olaya, M. Castellanos-Beltran, J. Pulecio *et al*., "Planarized Process for Single-Flux-Quantum Circuits with Self-Shunted Nb/NbxSi1−x/Nb Josephson Junctions," *IEEE Trans. Appl. Supercond*. **29**, 1101708 (2019).

[20] H. G. Ahmad, L. Di Palma, D. Massarotti *et al*., "Characterization of Lateral Junctions and Micro-SQUIDs Involving Magnetic Multilayers," *IEEE Trans. Appl. Supercond.* **33**, 1101205 (2023).

[21] S. K. Tolpygo *et al*., "Advanced fabrication processes for superconducting very large-scale integrated circuits," *IEEE Trans. Appl. Supercond*. **26**, 1100110, (2016).